\documentclass[aps, prb, preprint, showpacs, amssymb, amsmath, superscriptaddress]{revtex4-1}
\usepackage{mathrsfs}
\usepackage{inputenc}
\usepackage{bm}
\usepackage{times}
\usepackage{graphicx}
\usepackage[colorlinks,citecolor=blue,linkcolor=blue]{hyperref}
\usepackage{array}
\usepackage{float}
\usepackage{colortbl}
\usepackage{multirow}
\usepackage{tabularx}
\begin{document}
\title{Quantum spin Hall effect in two-dimensional transition-metal chalcogenides}

\makeatletter
\renewcommand*{\@fnsymbol}[1]{\ensuremath{\ifcase#1\or *\or \dag\or \dag \or
    \mathsection\or \mathparagraph\or \|\or **\or \dag\dag
    \or \dag\dag \else\@ctrerr\fi}}
\makeatother
\author{Xing Wang}
\affiliation{State Key Laboratory of Metastable Materials Science and Technology $\&$ Key Laboratory for Microstructural Material Physics of Hebei Province, School of Science, Yanshan University, Qinhuangdao 066004, China}
\affiliation{College of Science, Hebei North University, Zhangjiakou 07500, China}
\author{Wenhui Wan}
\affiliation{State Key Laboratory of Metastable Materials Science and Technology $\&$ Key Laboratory for Microstructural Material Physics of Hebei Province, School of Science, Yanshan University, Qinhuangdao 066004, China}
\author{Yanfeng Ge}
\affiliation{State Key Laboratory of Metastable Materials Science and Technology $\&$ Key Laboratory for Microstructural Material Physics of Hebei Province, School of Science, Yanshan University, Qinhuangdao 066004, China}
\author{Kai-Cheng Zhang}
\affiliation{College of Mathematics and Physics, Bohai University, Jinzhou 121013, China}
\author{Yong Liu}
\email{yongliu@ysu.edu.cn or ycliu@ysu.edu.cn}
\affiliation{State Key Laboratory of Metastable Materials Science and Technology $\&$ Key Laboratory for Microstructural Material Physics of Hebei Province, School of Science, Yanshan University, Qinhuangdao 066004, China}
\date{\today}


\begin{abstract}
  Based on first-principles calculations, we have found a family of two-dimensional (2D) transition-metal (TM) chalcogenides MX$_5$ (M = Zr, Hf and X = S, Se and Te) can host quantum spin Hall (QSH) effect. The molecular dynamics (MD) simulation indicate that they are all thermal-dynamically stable at room temperature, the largest band gap is 0.19 eV. We have investigated MX$_5$'s electronic properties and found their properties are very similar. The single-layer ZrX$_5$ are all gapless semimetals without consideration of spin-orbit coupling (SOC). The consideration of SOC will result in insulating phases with band gaps of 0.05 eV (direct), 0.18 eV (direct) and 0.13 eV (indirect) for ZrS$_5$, ZrSe$_5$ to ZrTe$_5$, respectively. The evolution of Wannier charge centers (WCC) and edge states confirm they are all QSH insulators. The mechanisms for QSH effect in ZrX$_5$ originate from the special nonsymmorphic space group features. In addition, the QSH state of ZrS$_5$ survives at a large range of strain as long as the interchain coupling is not strong enough to reverse the band ordering. The single-layer ZrS$_5$ will occur a topological insulator (TI)-to-semimetal (metal) or metal-to-semimetal transition under certain strain.  Monolayer MX$_5$ expand the TI materials based on TM chalcogenides and may open up a new way to fabricate novel low power spintronic devices at room temperature.
  \end{abstract}

\maketitle

\section*{I. Introduction}
Under the influence of graphene stripping from graphite, more and more two-dimensional (2D) materials are obtained by stripping block layered materials \cite{1,2,3}. These successfully prepared 2D materials have aroused great interest of scientists because of their excellent properties and great application prospects\cite{4,5,6,7,8}. Among them, 2D topological insulators (TIs) are of great significance to the study of condensed matter and materials science because of their fresh topological phases. The 2D TIs, also called quantum spin Hall (QSH) insulators, have fully spin-polarized gapless edge states into an insulating bulk. The QSH effect was first proposed by Kane and Mele in graphene \cite{9,10,11}, but its small energy gap of $10 ^{-3}$ meV makes it hard to be observed experimentally. The HgTe/CdTe quantum wells\cite{12,13} have been first realized in experiment, but it needs precisely controlled molecular beam epitaxy (MBE) growth at ultralow temperature.  In the passed years, extensive efforts have been devoted to search new QSH insulators, a growing number of compounds have been predicted to be 2D TIs \cite{14,15,16,17,18,19,20,21,22,23}. Due to the lack of suitable materials that are easy to prepare, stable in structure and large in band gap, the research of 2D TIs has been seriously hindered. Therefore search larger gap 2D TIs from the common used materials is indispensable for their ultimate realization.

At present, intensive research on transition metal (TM) based QSH insulators has greatly enriched the family of QSH insulators\cite{24,25,26}, showing band inversion is caused by strong electron interaction rather than spin-orbit coupling (SOC). TM atom-based TIs are rare and only several examples are reported. Among them, the MTe$_5$ (M = Zr, Hf) have attracted broad attention because of their topogical properties\cite{27,28,29,30,31,32,33,34,35}. Three-dimensional (3D)  MTe$_5$ is a layered crystal with weak interlayer coupling, which is comparable to graphite. The MTe$_5$ monolayer may be obtained  via the mechanical exfoliation from the 3D bulk phase as like producing graphene from graphite\cite{36,37}. The 2D crystals are predicted to be QSH insulators with the band gap about 0.1 eV\cite{27}. In addition, TM halide MX (M = Zr, Hf; X = Cl, Br and I) monolayer has to be found the 2D QSH insulators\cite{19}. Material design or atomic substitution may lead to more and better 2D TIs. Here we study TM chalcogenides MX$_5$ (M = Zr, Hf; X = S, Se and Te) and different TM chalcogenides may have different topological properties. What's more, we try to use the strain to increase the band gap of TM chalcogenides.

In this work, we have investigated the structural stability and electronic structure of single-layer MX$_5$ by using first-principles calculations. The single-layer ZrX$_5$ and HfX$_5$ have very similar properties, they are all QSH insulators with the largest band gap of 0.19 eV.  Then we take ZrX$_5$ as an example, they are all gapless semimetals without consideration of SOC. The consideration of SOC will result in QSH insulators with band gaps of 0.05 eV (direct), 0.18 eV (direct) and 0.13 eV (indirect) for ZrS$_5$, ZrSe$_5$ to ZrTe$_5$, respectively. The topological invariant Z$_2$=1 and edge states confirm the nontrivial topological nature of these materials. The mechanism of QSH effect in ZrX$_5$ originates from the special nonsymmorphic space group features. In addition, the QSH state of ZrS$_5$ survives over a very large strain range until the interchain coupling is strong enough to reverse the band ordering.

\section*{II. COMPUTATIONAL DETAILS}
The Vienna ab initio Simulation Package (VASP)\cite{38,39} is used to study the structural and electronic properties of MX$_5$. The exchange correlation interaction is treated within the generalized gradient approximation (GGA)\cite{40} of Perdew, Burke and Ernzerhof (PBE)\cite{41}. The Brillouin zone (BZ) is integrated with $13\times4\times1$ $\Gamma$-centered Monkhorst-Pack grid\cite{42}, the plane-wave cutoff energy is set to 500 eV with the energy precision of 10$^{-5}$ eV.  A more than 20 {\AA} vacuum slab along \textit{z}-direction is set to avoid interactions between neighboring slabs. The atoms in the unit cell are relaxed until the force on each atom is less than 0.01 eV/{\AA}. Since PBE methods often underestimate the band gap, the hybrid functional HSE06\cite{43} is used to check the results and the band gaps are very similar.  Further, the local orbital basis suite towards electronic-structure reconstruction (LOBSTER)\cite{44,45,46} has been used to extract the chemical-bonding information for ZrX$_5$. The tight binding matrix elements are calculated by projecting the Bloch states onto maximally localized Wannier functions (MLWFs)\cite{47,48} using the VASPWANNIER90 interface. The MLWFs are derived from M's d and X's p orbitals by using the Wannier90 code\cite{47,48}. The WannierTools\cite{49} is used to analyse topological properties after successful constructions of the MLWFs.

\section*{III. RESULTS AND DISCUSSION}

\begin{figure}[htp!]
\centerline{\includegraphics[width=0.8\textwidth]{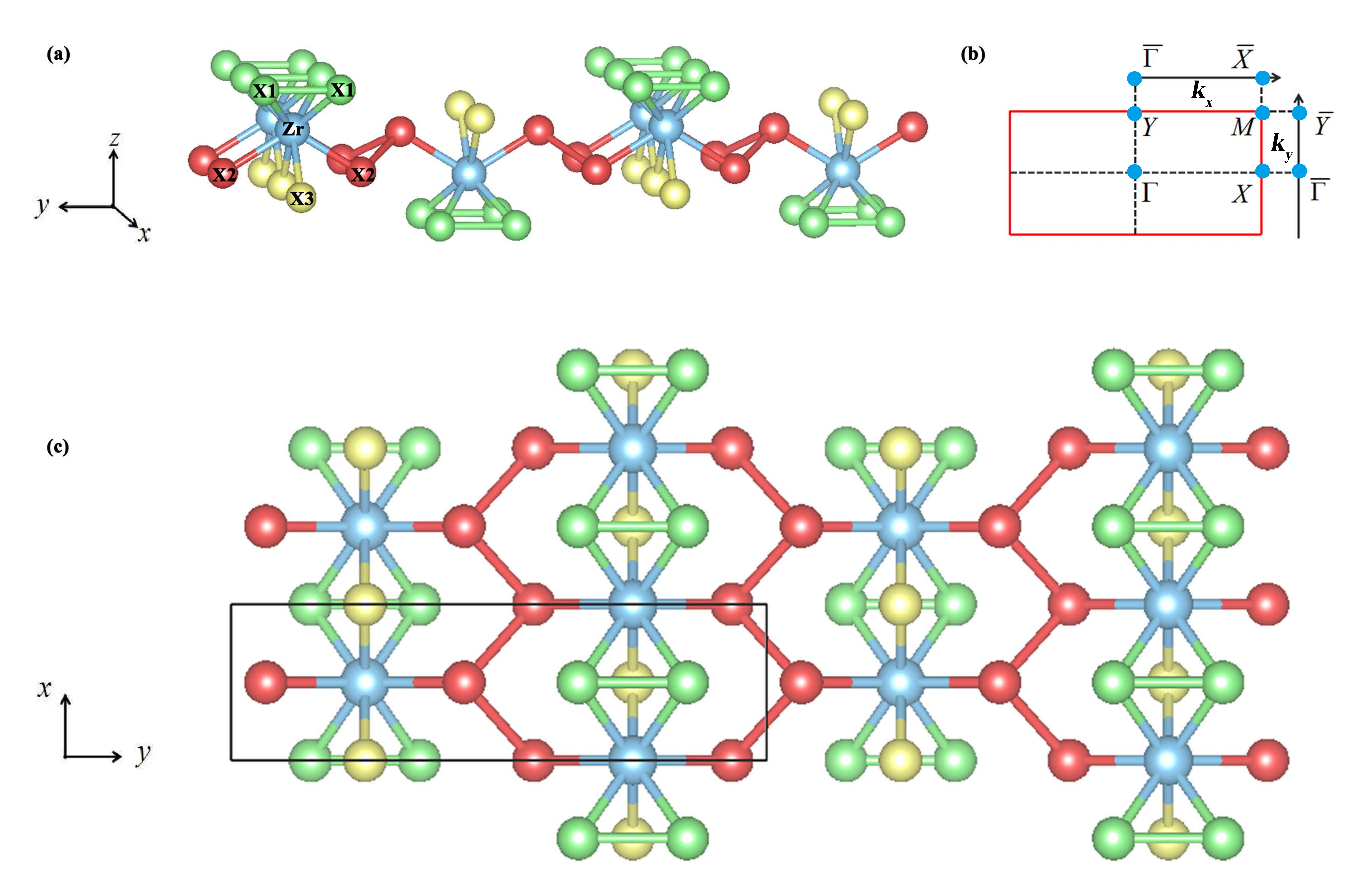}}
\caption{(Color online) The  side view (a) and  top view (c) of relaxed single-layer MX$_5$. Blue balls are M atoms and the other color balls are X atoms. The primitive cell is shown in black rectangle. (b) 2D and projected edge first BZ with high symmetry points (blue dots). }
\end{figure}

The MX$_5$ monolayer has the orthorhombic structure with Pmmn ($D_{2h}^{13}$) nonsymmorphic space group shown in Fig. 1 (a). The trigonal prismatic chains X1-X3-X1 oriented along the x axis, and these prismatic chains are linked via parallel zigzag chains of X2 atoms along the y axis to form a 2D sheet of MX$_5$ in the x-y plane. We choose the origin of the coordinate system located on Zr site, so the inversion center is located at (0.25, 0.25). The two X1 (or X2) atoms in the same chain have mirror symmetry, while the two X3 (or X2) atoms in the adjacent prism (or same zigzag) chain have inversion symmetry. The optimized lattice constants of single-layer MX$_5$ are summarized in Table I. The relaxed lattice constants ($\textit{a}$= 4.047 {\AA}, $\textit{b}$= 13.860 {\AA}) of ZrTe$_5$ are nearly same to the reference value ( $\textit{a}$= 4.036 {\AA}, $\textit{b}$= 13.843 {\AA})\cite{27}. The lattice constants have increase trend with increasing atomic radius of X atom in ZrX$_5$ and HfX$_5$ monolayer. While fixed chalcogenide atoms, the atomic radius of Zr and Hf are affected by lanthanide contraction effect\cite{50}, so the lattice constants show opposite variation trends, similar to the ZrCl and HfCl\cite{19}. The single-layer ZrX$_5$ and HfX$_5$ have very similar properties, hereafter, we take ZrX$_5$ as an example.

First, we use the crystal orbital Hamilton population (COHP) method to extract the chemical-bonding information, shown in Table II. The bond lengths increase from ZrS$_5$, ZrSe$_5$ to ZrTe$_5$, and it makes sense because the atomic radius increases from S, Se to Te. The Fermi levels in Fig. 2 (a) and supplemental material (SM) Fig. S1 lie in the COHP curve between the bonding and antibonding regions. Fousing on the plots below the Fermi level, the majority of the bonding interactions resides the atoms Zr-X, the majority of the antibonding interactions reside the atoms X-X. But there are not many populated antibonding states, which doesn't lead to an enormous internal stress, so the materials are relatively stable. The integrated COHP (ICOHP) values increase from ZrS$_5$, ZrSe$_5$ to ZrTe$_5$, which indicates the material stability decreases in turn. In addition, the binding energy of -2.305, -1.985 and -1.613 eV/atom indicates the materials are all stable.  Moreover, we examine thermal stability of ZrX$_5$ by performing ab initio molecular dynamics (MD) simulation. After heating at 300 K for 10ps with a time step of 2fs, it is found that the average of the total potential energy remains constant throughout the simulation time, see Fig. 2 (b) and Fig. S2. These materials do not undergo structural reconstruction or damage. These results indicate clearly the materials remain thermal dynamically stable at room temperature.

Then the electron localization function (ELF)\cite{51} are used to describe and visualize chemical bonds in ZrX$_5$ monolayer. The result for single-layer  ZrS$_5$ is illustrated in Fig. 2 (c) for the ELF = 0.88 isosurface. The greater value of S-S bonding suggests anti-bonding character, while the Zr-S bonding indicates the highly ionic nature. Charge transfer of Zr-S bonding is studied by difference charge density diagram, as shown in Fig. 2 (d). The charge  accumulation (loss) is represented by red (yellow) region, the major charge transfer is from Zr atom to S atom, which is consistent with ionic bonding. The ZrSe$_5$ and ZrTe$_5$ have similar characters, presented in Fig. S3 and S4.

\begin{table*}[!htbp]
\centering
\caption{The lattice constants \textit{a}, \textit{b}, global band gap E$_g$, band gap at $\Gamma$ point E$_{\Gamma}$ and Z$_2$ invariant of single-layer MX$_5$.}
\begin{tabular}{c|c|c|c|c|c} 
\hline
\hline
 Material&$\textit{a}$({\AA})&$\textit{b}$({\AA})&E$_g$(eV)&E$_{\Gamma}$(eV)&Z$_2$\\
\hline
ZrS$_5$&3.529&11.679&0.050&0.835&1\\
ZrSe$_5$&3.731&12.487&0.181&0.509&1\\
ZrTe$_5$&4.047&13.860&0.129&0.400&1\\
\hline
HfS$_5$&3.499&11.637&0.049&0.757&1\\
HfSe$_5$&3.703&12.457&0.187&0.474&1\\
HfTe$_5$&4.020&13.853&0.106&0.383&1\\
\hline
\hline
\end{tabular}
\end{table*}

\begin{table*}[!htbp]
\centering
\caption{The bond lengths and ICOHP values of ZrTe$_5$, ZrSe$_5$ and ZrS$_5$, respectively.}
\begin{tabular}{c|c|c|c|c|c|c} 
\hline
\hline
 2D&$\textit{d$_{Zr-X1}$}$&ICOHP$_{Zr-X1}$&$\textit{d$_{Zr-X2}$}$&ICOHP$_{Zr-X2}$&$\textit{d$_{Zr-X3}$}$&ICOHP$_{Zr-X3}$\\
 Materials&({\AA})&(eV/bond)&({\AA})&(eV/bond)&({\AA})&(eV/bond)\\
\hline
ZrS$_5$&2.56&-3.03&2.67&-2.40&2.58&-2.74\\
ZrSe$_5$&2.74&-2.54&2.78&-2.35&2.75&-2.54\\
ZrTe$_5$&2.97&-2.04&2.99&-2.08&3.01&-2.21\\
\hline
\hline
\end{tabular}
\end{table*}

\begin{figure}[htp!]
\centerline{\includegraphics[width=1.0\textwidth]{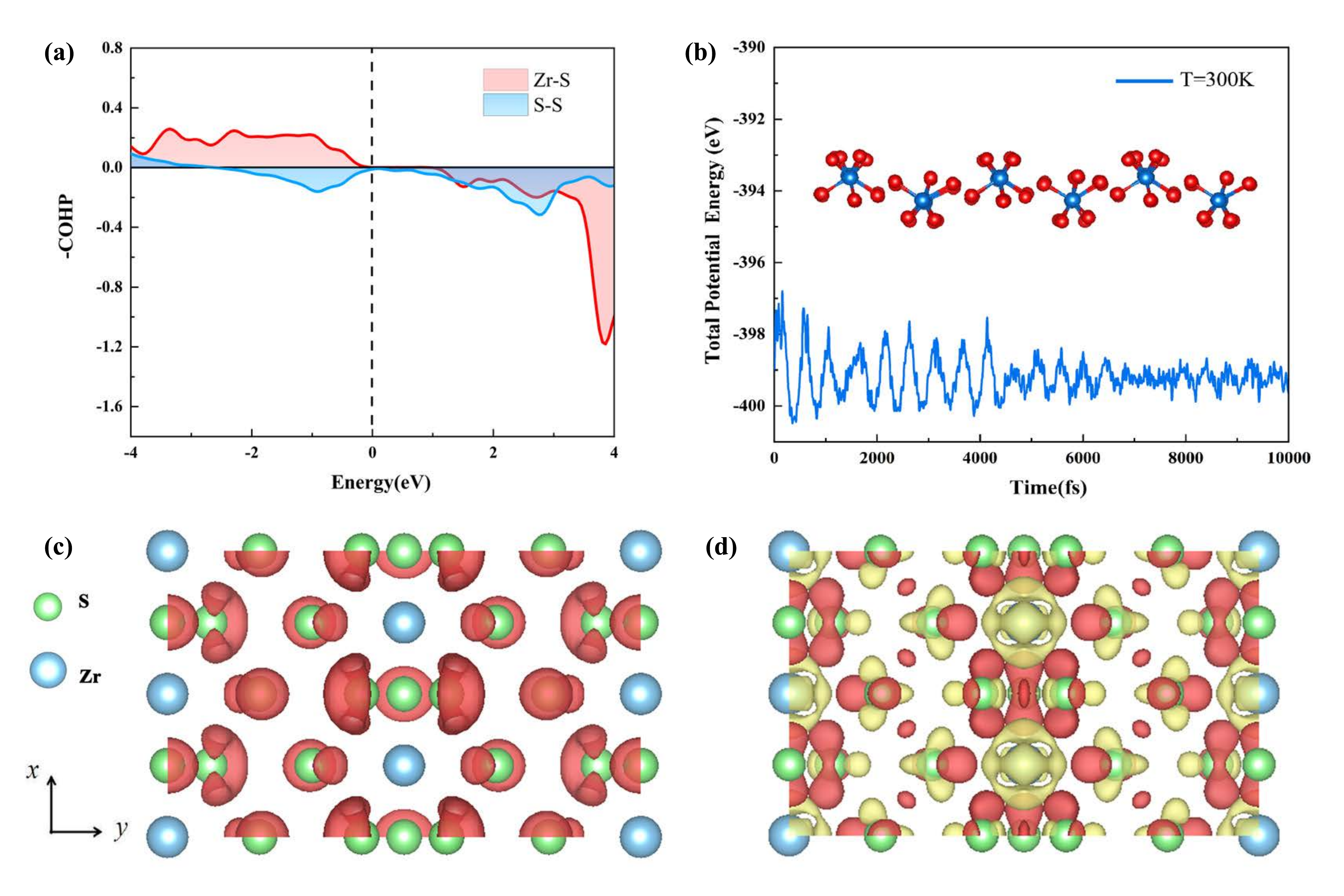}}
\caption{(Color online) The crystal orbital Hamilton population (COHP) curves, molecular dynamics (MD) simulation, electron localization function (ELF) and difference charge density of single-layer ZrS$_5$. (a) The COHP curves. The Fermi levels are marked by the vertical dashed lines. (b) Snapshots of atomic configurations at the end of MD simulation and total potential energy fluctuations observed at 300K. (c) Structure plot of ELF. Isosurface corresponding to ELF value of 0.88. (d) Difference charge density. The red (yellow) isosurface plots correspond to the charge density accumulation (depletion). }
\end{figure}

To further understand of their electronic properties, we have calculated the total density of states (DOS) and projected density of states (PDOS) with consideration of SOC for ZrX$_5$, shown in Fig. 3 and Fig. S5. The small band gaps near the Fermi levels indicate they are all semiconductors, so we take the ZrS$_5$ as an example.  The PDOS clearly show that around Fermi level the S covalently bonded p states are dominant. The $d$ orbitals of Zr are mainly located at the level of  4eV above the Fermi level, leading to the nearly ionic states. These results are consistent with the previous discussion.

 \begin{figure}[htp!]
\centerline{\includegraphics[width=1.0\textwidth]{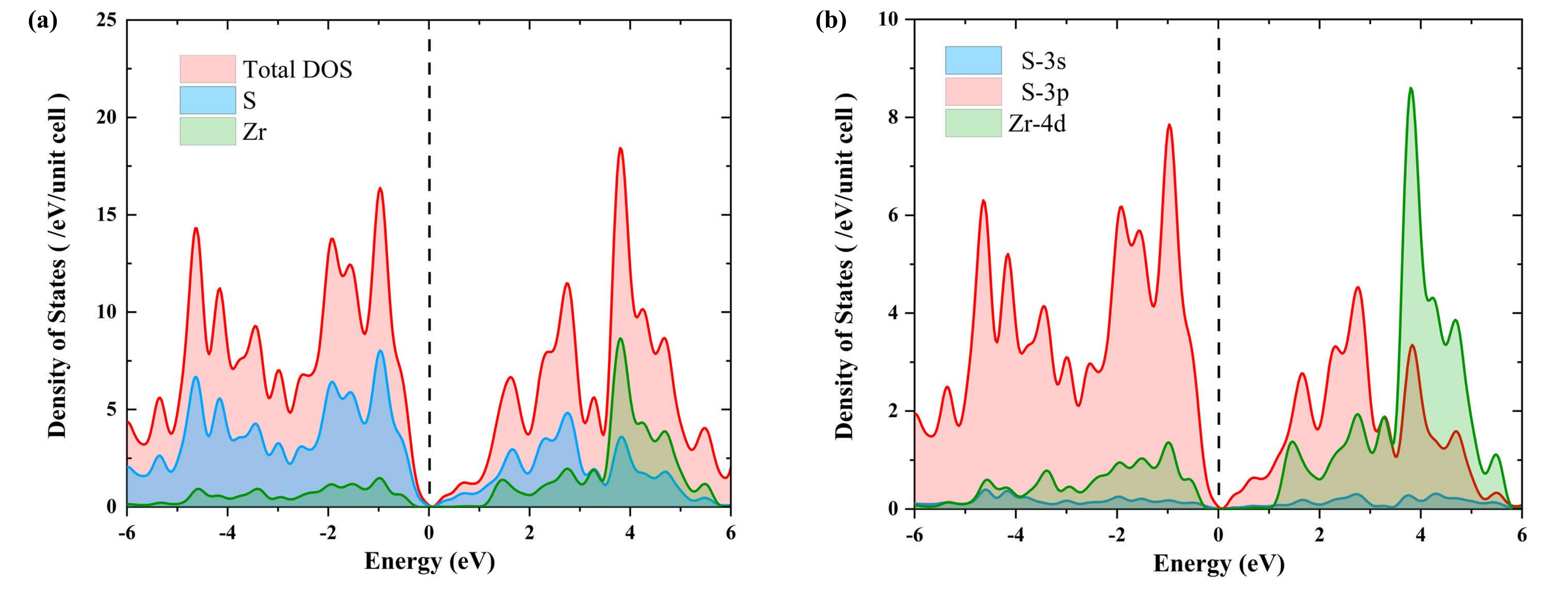}}
\caption{(Color online) The total density of states (DOS) (a) and projected density of states (PDOS) (b) with consideration of SOC for single-layer ZrS$_5$. The Fermi levels are set to zero and marked by vertical dashed lines.}
\end{figure}

\begin{figure}[htp!]
\centerline{\includegraphics[width=1.0\textwidth]{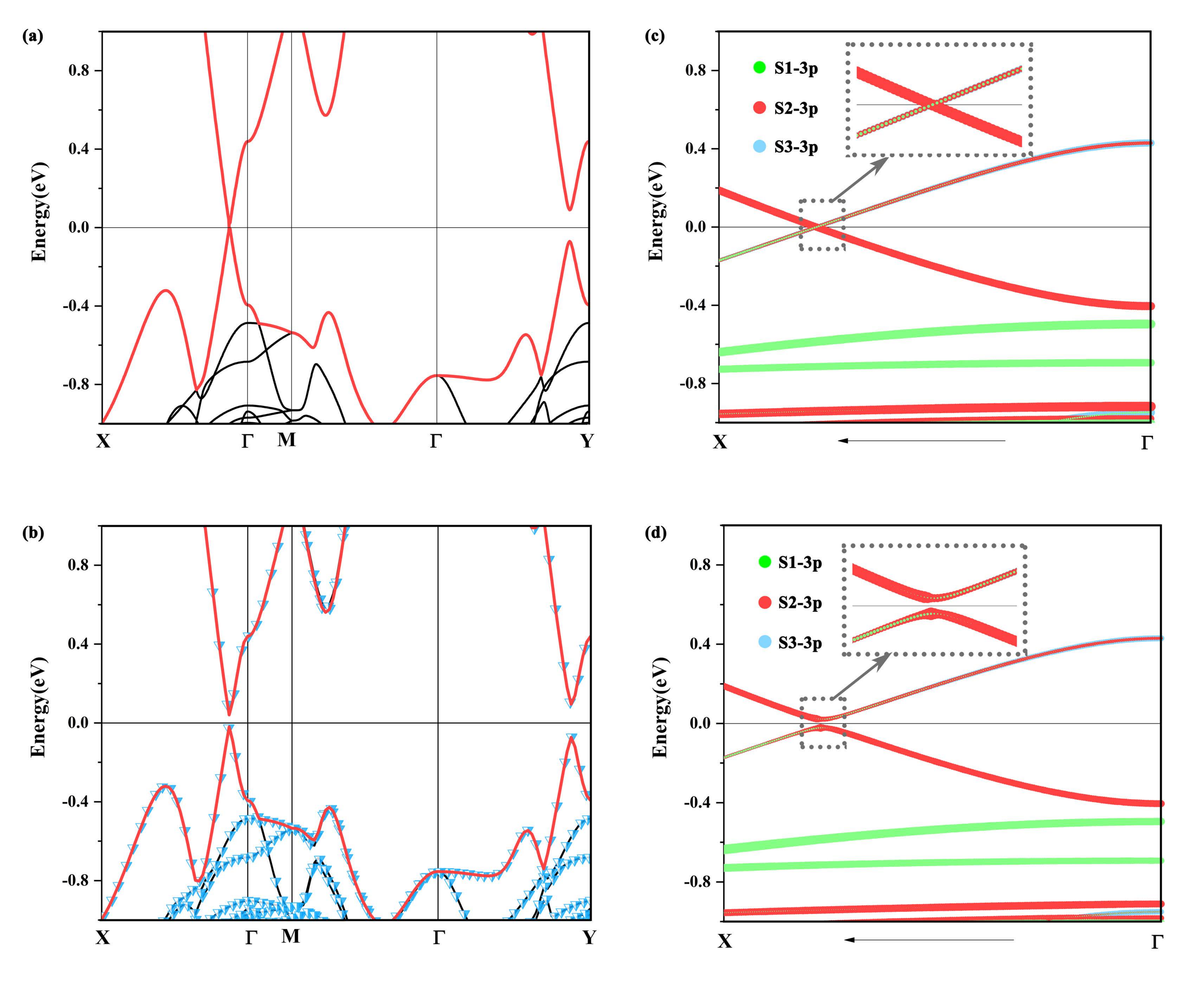}}
\caption{(Color online) The band structures of single-layer ZrS$_5$ for (a, c) without and (b, d) with SOC case. The blue triangles indicate the band structures diagrams calculated by Wannier90. The enlarged orbitals-resolved band structures are shown in insets. The green, red and blue circles represent the weights of the S1-3p, S2-3p and S3-3p orbitals character, respectively. The Fermi levels are set to be zero.}
\end{figure}

\begin{figure}[htp!]
\centerline{\includegraphics[width=1.0\textwidth]{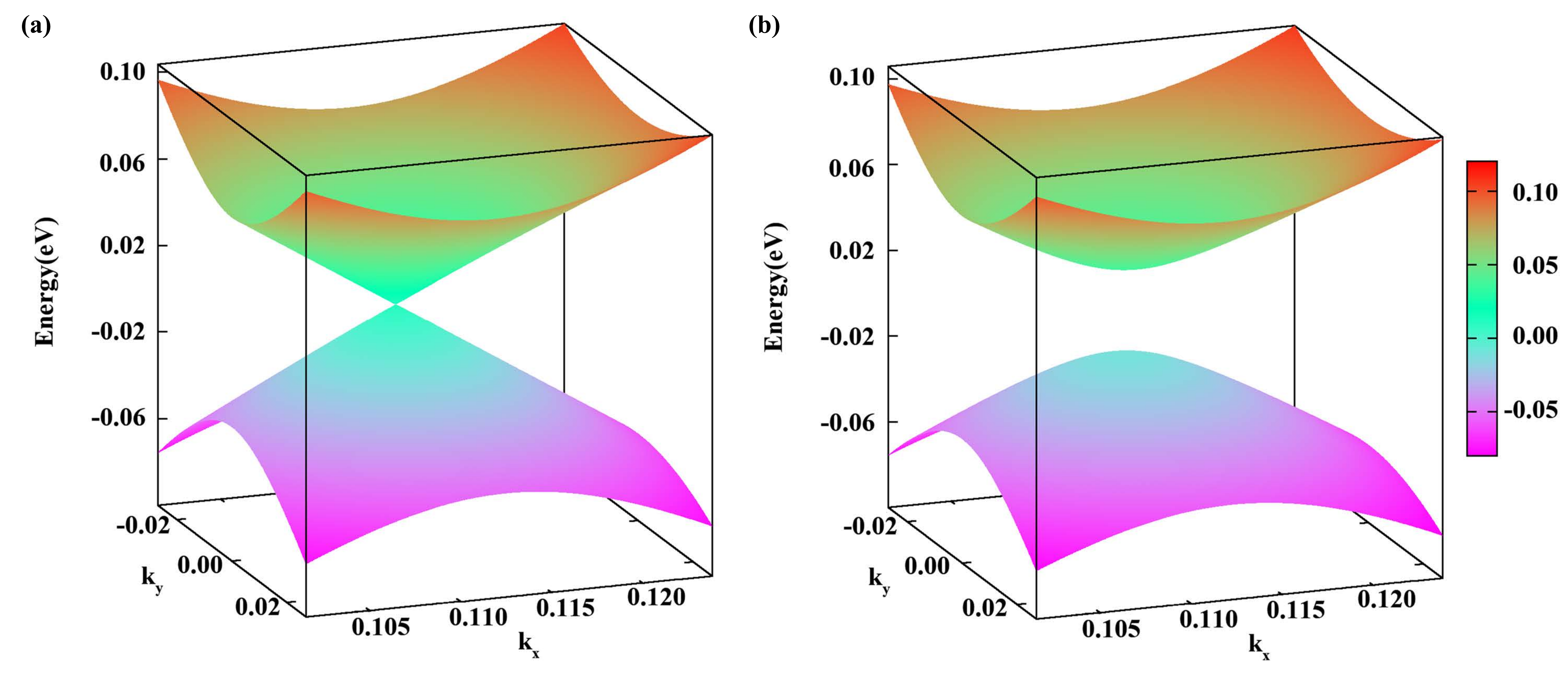}}
\caption{(Color online) The 3D band structures of single-layer ZrS$_5$ (a) without and (b) with SOC, respectively. The Fermi levels are set to zero.}
\end{figure}

The calculated band structures for a single-layer ZrX$_5$ are displayed in Fig. 4 and Fig. S6. They are all  gapless semimetals without consideration of SOC, the conduction band minimum (CBM) and valence band maximum (VBM) touch each other at the $\Gamma$-X direction.  In Fig. 5 (a), we plot the ZrS$_5$'s band dispersion around this crossing point, which demonstrates that the Dirac point is isolated and has linear dispersion.  The consideration of SOC will result in insulator phase and the large band gaps at $\Gamma$ point decrease from ZrS$_5$, ZrSe$_5$  to ZrTe$_5$, shown in Table I. The global band gaps of 0.05 eV (direct), 0.18 eV (direct) and 0.13 eV (indirect) for ZrS$_5$, ZrSe$_5$ to ZrTe$_5$, respectively. The ZrS$_5$ and ZrSe$_5$ are more suitable than ZrTe$_5$ as materials for making photoelectric devices.  We can clearly see a small gap will be opened at the original Dirac point and the low-energy electrons become 2D massive Dirac fermions in Fig. 5 (b).  The evolution lines of Wannier centers in Fig. 6 (a) and Fig. S7 show they are all nontrivial QSH insulators with Z$_2$=1 and the important character of helical edge states also appear, as shown in Fig. 6 (b, c) and  Fig. S8. The nontrivial Z$_2$ invariants guarantee that the edge bands always cut the Fermi level an odd number of times. The helical edge states form bands crossing linearly at the $\overline{\Gamma}$ point for the \textit{x} axis edge in the bulk gap. Each edge has a single pair of helical edge states for these systems. The two counter-propagating edge states display opposite spin-polarizations is the typical feature of a QSH phase. Helical edge states are very important in electronics and spintronics because of their topological robustness to scattering.  As an important characterization related to application, the Fermi velocity of helical edge states are about $1.0\times 10^4$ m/s, $2.3\times 10^5$ m/s and $2.0\times 10^5$ m/s for ZrS$_5$, ZrSe$_5$ to ZrTe$_5$, respectively. The values are a little smaller than these for stanene ($4.4\times 10^5$ m/s), fluorinated stanene ($6.8\times 10^5$ m/s), and HgTe quantum well ($5.5\times 10^5$ m/s)\cite{15}. It can be expected that the ZrSe$_5$ and ZrTe$_5$ are more suitable than ZrS$_5$ as candidate materials for high-speed devices. For the edge along the \textit{y} axis, the symmetric edge structure leads to two Dirac cones located at opposite $\overline{Y}$ points.  The topological nature of monolayer ZrX$_5$ is further confirmed by the nontrivial metallic edge states.

\begin{figure}[H]
   \centering
  \includegraphics[scale=0.13]{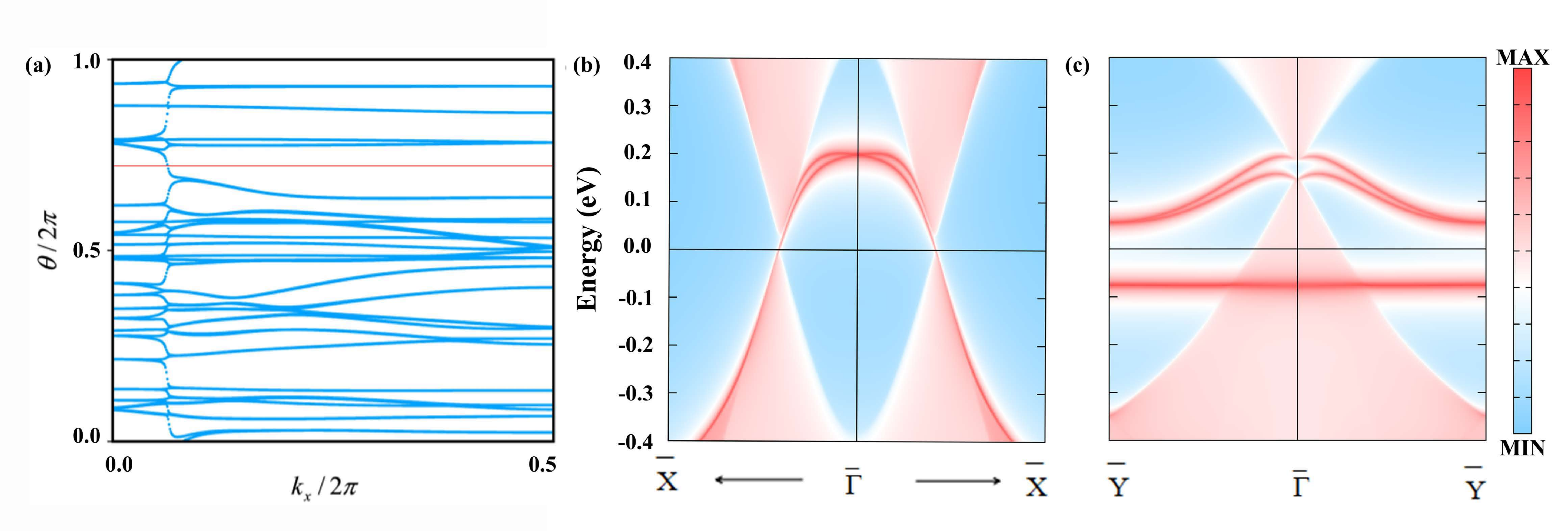}
\caption{(Color online) The evolution of Wannier charge centers (WCC) and edge states of single-layer ZrS$_5$. (a) The evolution of WCC along k$_x$. The evolution blue lines cross the arbitrary reference red line one time yielding Z$_2$=1.  The edge state for (b) \textit{x} edge and (c) \textit{y} edge. }
\end{figure}

\begin{figure}[htp!]
\centerline{\includegraphics[width=1.0\textwidth]{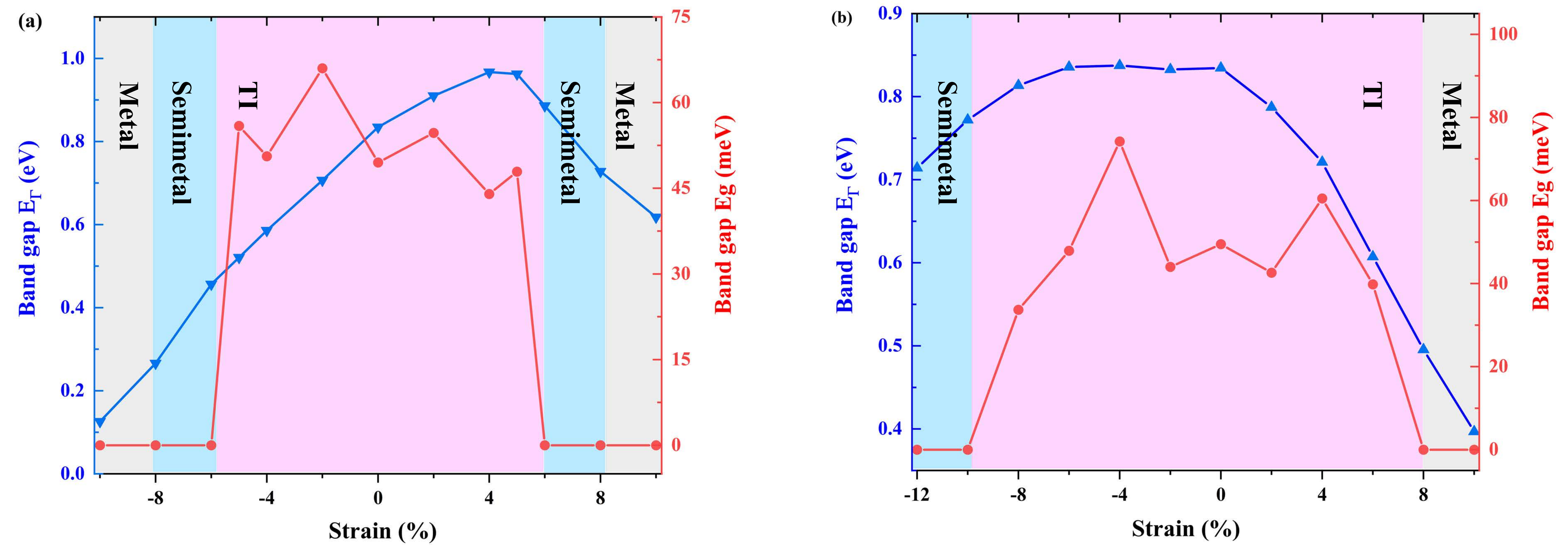}}
\caption{(Color online)The variation of band gap as a function of uniaxial strain along the (a) [100] and (b) [010] direction of single-layer ZrS$_5$. E$_g$ and E$_\Gamma$ represent the globe band gap and direct band gap at the $\Gamma$ point, respectively.  The nontrivial Z$_2$ topology survives as long as the globe band gap remains positive.}
\end{figure}

To get a physical understanding of the topological nature, we present the orbitals-resolved band structures in Fig. 4 (c, d) and SM Fig. S6, it can be seen the bands around the Fermi level are mainly derived from X1-p and X2-p orbitals for ZrX$_5$, so they have the same band inversion mechanism as ZrTe$_5$\cite{27}.  The mechanism for QSH effect in ZrX$_5$ originates from the special nonsymmorphic space group features, X1 state has odd parity lead to the total parity of the occupied states is negative, which leads to the QSH state.

Further more, we study electronic properties of single-layer ZrS$_5$ under different strain to explore the possible phase transition. For the uniaxial strain along the [100] and [010] directions, the variation of band gap as a function of strain is presented Fig. 8. Strain has little effect on the global band gap E$_g$, but greatly affects band gap at the $\Gamma$ point band gap E$_\Gamma$. For the [100] uniaxial strain, the E$_{\Gamma}$ increases first then decreases from -10 \% to 10\% and reaches a maximum value of 0.97 eV at 4\% tensile strain. The E$_g$ varies with strain and reaches a maximum value of 66.0 meV at 2\% compressive strain. When the compressive (or tensile) strain is more than 6\%, the band structure produces a TI-to-metal transition. When the compressive (or tensile) strain is more than 8\%, the band structure produces a metal-to-semimetal transition.

For the [010] uniaxial strain, the E$_{\Gamma}$ also increases first then decreases from -12\% to 10\%, reaches a maximum value of 0.84 eV at 4\% compressive strain. However, the E$_g$ value changes very little with strain, reaches a maximum value of 74.2 meV at 4\% compressive strain. When the compressive strain is more than 10\%, the band structure produces a TI-to-semimetal transition. When the tensile strain is more than 8\%, the band structure produces a TI-to-metal transition. In a word, the nontrivial topological phases survive over a wide strain from -6\% to 6\% along [100] direction or -10\% to 8\% along [010] direction, as long as the interchain coupling is not strong enough to reverse the band ordering near the Fermi energy, the QSH state should be stable, such robust topology gainst lattice deformation may make it easier for experimental realization and characterization on different substrate.

Finally, we have analyzed the electronic and topological properties for HfX$_5$ monolayer and draw the same conclusions as ZrX$_5$. HfX$_5$ mnolayer materials also possess QSH states with Z$_2$ = 1, remain thermal dynamically stable at room temperature, and their  band structures and band-gap sizes are very similar to ZrX$_5$'s, see Table II, Fig. S9 and Fig. S10.

\section*{IV. Conclusion}
In summary, we have found a family of 2D MX$_5$ can host QSH effect. The MD simulation indicates they are all thermal-dynamically stable at room temperature, they are all QSH insulators with the largest band gap is 0.19 eV. The single-layer ZrX$_5$ and HfX$_5$ have very similar properties.  The evolution of WCC and edge states verifies the nontrivial topological nature of these materials. The Dirac points of these materials have high velocities, so they can be used as candidates for high-speed electronic devices materials. The mechanism for QSH effect in ZrX$_5$ originates from the special nonsymmorphic space group features, X1 state has odd parity lead to the total parity of the occupied states is negative, which leads to the QSH state. In addition, the QSH state of ZrS$_5$ survives at a large range of strain as long as the interchain coupling is not strong enough to reverse the band ordering. The band structure of single-layer ZrS$_5$ can produce transition among TI, semimetal and metal under certain strain. These simple monolayer ZrX$_5$ without chemical decoration, which are very favorable for the future experimental implementation via simple exfoliation from its 3D system, making them highly adaptable to various environments. These significant results may promote the further study of QSH insulators based on TM chalcogenide.

\begin{acknowledgments}
This work was supported by National Natural Science Foundation of China (No. 11904312 and 11904313), the Project of Department of Education of Hebei Province, China(No. BJ2020015), and the Natural Science Foundation of Hebei Province (No. A2019203507 and A2020203027). K.C. Zhang acknowledges the fund support from LiaoNing Revitalization Talents Program (No. XLYC2007120). The authors thank the High Performance Computing Center of Yanshan University.
\end{acknowledgments}


\end{document}